\documentclass[a4paper,fleqn,usenatbib]{mnras}

\usepackage[T1]{fontenc}
\usepackage{ae,aecompl}

\usepackage{graphicx}	
\usepackage{amsmath}	
\usepackage{amssymb}	

\newcommand{\be}{\begin{equation}}
\newcommand{\ee}{\end{equation}}
\def\ergs{{\rm\,erg\,s^{-1}}}
\newcommand{\msun}{{M}_{\sun}}

\title[Constraint on the M87 black hole spin]{Constraint on the black-hole spin of M87 from the accretion-jet model}

\author[Feng and Wu]{
Jianchao Feng,$^{1}$
and Qingwen Wu,$^{1}$\thanks{E-mail: qwwu@hust.edu.cn}
\\
$^{1}$School of Physics, Huazhong University of Science and Technology,
 Wuhan 430074, China\\
}

\date{Accepted XXX. Received YYY; in original form ZZZ}

\pubyear{2016}

\begin{document}
\label{firstpage}
\pagerange{\pageref{firstpage}--\pageref{lastpage}}
\maketitle

\begin{abstract}
The millimeter bump, as found in high-resolution multi-waveband observations of M87, most possibly comes from the synchrotron emission of thermal electrons in advection dominated accretion flow(ADAF). It is possible to constrain the accretion rate near the horizon if both the nuclear millimeter emission and its polarization are produced by the hot plasma in the accretion flow. The jet power of M87 has been extensively explored, which is around $8_{\rm -3}^{+7}\times10^{42}$ {\rm erg/s} based on the analysis of the X-ray cavity. The black hole(BH) spin can be estimated if the jet power and the accretion rate near the horizon are known. We model the multi-wavelength spectral energy distribution (SED) of M87 with a coupled ADAF-jet model surrounding a Kerr BH, where the full set of relativistic hydrodynamical equations of the ADAF are solved. The hybrid jet formation model, as a variant of Blandford-Znajek model, is used to model the jet power. We find that the SMBH should be fast rotating with a dimensionless spin parameter $a_{*}\simeq0.98_{\rm -0.02}^{+0.012}$.
\end{abstract}


\section{Introduction}
It is now widely believed that the center of each galaxy harbors a super-massive black hole (SMBH) with a mass of the order of $10^6-10^9 \msun$.
The surrounding material close the BH can be captured and form an accretion disk, where the structure and radiation of the accretion disk have been extensively studied in last several decades. The bright state of X-ray binaries (XRBs) and luminous active galactic nuclei (AGNs, e.g. QSOs) are believed to be powered by a cold, optically thick, geometrically thin standard accretion disc \citep[SSD, e.g.,][]{ss73} that accompanied with some fraction of hot optically thin corona above and below the disc. The cold disc may transit to the hot, optically thin, geometrically thick ADAF \citep[e.g.,][]{ny95,ab95,yn14} when the accretion rate is less than a critical value (e.g., low-state of XRBs and faint AGNs). The possible transition of the accretion modes is supported by the observational features of XRBs and AGNs, where the bolometric Eddington ratio and X-ray spectral index show anti- or positive correlation when the Eddington ratio is less or larger than a critical value of $\sim 1\%$ \citep[e.g.,][]{wu08,wang04,sh08,con09,gu09}.

M87 is the one of the closest AGNs \citep[$D=16.7\pm0.6$ Mpc][]{jord05,blak09} that observed with a large-scale relativistic jet. The mass of the central SMBH is $3-6.6\times10^9 \msun$ \citep{macc97,gebh11,wals13}. Due to its proximity and the large BH mass, the SMBH in M87 is one of the largest BHs on the sky, with putative event horizons subtending 38 $\mu$as, which is slightly less than but more or less similar to that of the Galactic center BH (Sgr A*) with event horizons subtending 53 $\mu$as \citep[e.g.,][]{rd15}. Comparing no evident jet in Sgr A*, the radio core of M87 have been resolved into a clear jet structure, which can be used as a laboratory for exploring various theoretical models of accretion-jet physics. The bolometric luminosity of the core is estimated to be $L_{\rm bol}\sim2.7\times10^{42} \ergs \sim 3.6\times10^{-6} L_{\rm Edd}$ \citep[$L_{\rm Edd}\ergs$ is Eddington luminosity,][]{pri16}, which is several orders of magnitude less than those of Seyferts and quasars. The quite low Eddington ratio in M 87 suggests that it most possibly accretes through an ADAF \citep[see][for a recent review and references therein]{yn14}. The strong jet has been observed in the multi-waveband from the radio to the $\gamma$-ray~\citep[e.g.,][]{rei89,jun99,per05,har06,ly07,kov07,doe12,ak15,had16}. The cavities seen in the X-ray emission of Virgo cluster provides a direct measurement of the nonradiative energy output via jets from the central BH of the M87, where the jet power is $8_{\rm -3}^{+7}\times10^{42}$ {\rm erg/s} \citep[e.g.,][]{raf06,russ13}.

The physical mechanism for the formation of the relativistic jet is still unknown. The currently most favored jet formation mechanisms include the Blandford-Znajek (BZ) process \citep[]{bz77} and the Blandford-Payne (BP) process \citep[]{bp82}. In the BZ process, the energy extraction is purely electromagnetic and is directly coupled to the spin of the BH. The possible spin paradigm was also supported from observations, where \citet{ny12} found that the jet radio luminosity (which assumed to be associated to jet power) scales as the square of the BH spin by using five XRBs with the BH spin measurements from the thermal continuum-fitting method \citep[but see also][]{russ13a}. In the BP process, however, the particle acceleration is done through a magnetic sling-shot mechanism with the field lines anchored in the rotating accretion flow \citep[e.g., see also][]{li12,cao12,cao14}. The recent magnetohydrodynamic (MHD) simulations showed that both BH spin and accretion process may play important roles in jet formation \citep[e.g.,][]{mck04,hir04,vill05,haw06}. The so-called hybrid model, as a variant of the BZ model, was proposed by \citet{mei99}, which combined the BZ and BP effects through the large-scale magnetic fields threading the accretion disk outside the ergosphere and the rotating plasma within the ergosphere.

Several methods have been proposed to estimate the BH spin, including the study of the thermal spectrum of the standard thin disk\citep[e.g.,][]{zh97,gou11,you16}, the analysis of the reflection spectrum of the $K\alpha$ iron line \citep[e.g.,][]{fab89,gou11}, the measurement of jet formation efficiency \citep[e.g.,][]{tch10,tch11,tch12a,tch12b} and the precession of disk-jet \citep[e.g. Bardeen-Petterson effect,][]{bp75,wu13,lei13}.  However, it is still difficult to estimate the BH spin of M87, due to there is no standard disk, no broad $K\alpha$ line and/or evident jet precession effect etc.. \citet{li09} and \citet{wang08} proposed that the TeV photons could be used as a diagnostic for estimating black hole spin, where the dimensionless BH spin $a_{*}\gtrsim0.8$ for M87 based on the transparent radius of the high-energy photons detected by H.E.S.S.. In our former paper, we constrained the accretion rate near the BH horizon for M87 through the recent the Faraday rotation measure and Millimeter/submillimeter high-resolution observations \citep[]{feng16} \citep[see also][]{li16}. Therefore, it is possible to constrain the BH spin from the jet formation model with the estimated accretion rate near the BH horizon and the jet power. The estimation of M87 BH spin will help us to further explore the accretion-jet physics in strong gravitational field near the horizon in near future (e.g., BH shadow etc.). In Section 2, we present the ADAF-jet model, results and discussions are given in Section 3 and 4, respectively. We summarize the main result in Section 5. Throughout this work, we adopt a BH mass of $6.6\times10^9 \msun$, a jet power of $8_{\rm -3}^{+7}\times10^{42}$ {\rm erg/s} and a distance of 16.7 Mpc.

\section{Method}
\subsection{ADAF-jet model}
We adopt a coupled ADAF-jet model surrounding a Kerr BH to constrain the BH spin in M87, where both the multibands emission and jet power are considered. We simply introduce the accretion and jet model as below, and more details can be found in \citet{wu13}, \citet{feng16} and references therein.

The global structure of the ADAF in general relativistic frame is solved numerically, where the ion and electron temperature, density, angular momentum, radial velocity at each radius can be obtained. The accretion rate is $\dot{M}=\dot{M}_{\rm out}(R/R_{\rm out})^{\it s}$, where the possible wind is considered ($s$ is the wind parameter) and $\dot{M}_{\rm out}$ is the accretion rate at the outer radius, $R_{\rm out}$. We simply set the accretion rate at outer boundary equal to the Bondi accretion rate at Bondi radius ($\dot{M}_{\rm out}=\dot{M}_{\rm B}=0.2\ \msun \rm yr^{-1}$). The global structure of the ADAF can be calculated if the parameters $a_{*}$, $\alpha$, $\beta$, and $\delta$ are given, where $a_{*}=Jc/GM_{\rm BH}^2$ is the dimensionless BH spin, $\alpha$ is viscosity parameter, $\beta$ is the ratio of gas to total pressure (sum of gas and magnetic pressure), and $\delta$ describes the fraction of the turbulent dissipation that directly heats the electrons in the flow \citep*[see][for more details]{man00}. For $\alpha$ and $\delta$, we adopt typical values of 0.3 and 0.1 as widely used in ADAF models, respectively. The value of $\beta$ is typically $\sim0.5-0.9$ \citep[][]{yn14}, where $\beta=0.5$ correspond to the equipartition between magnetic energy and thermal energy. We keep $s$ as a free parameter, which can be constrained in SED fitting if other parameters are fixed. The radiation processes of synchrotron, bremsstrahlung, and Compton scattering are considered consistently in our calculations.

For the jet model, both the power and radiation are considered. Recent numerical simulations show that the coupling between the accretion flow and the magnetic field is an essential element of jet production \citep[e.g.,][]{vill05,haw06,tch10,tch11,tch12a,tch12b}, where the jet is weak and still exist even the BH spin $a_{*}=0$. This is different from that the pure BZ model. In this work, we adopt the hybrid jet model that proposed by \citet{mei01}, where the jet extract the BH rotational energy indirectly through the large-scale magnetic field that anchored to the accretion disc outside the ergosphere as well as rotating plasma within the ergosphere. The field amplifications by both differential rotation of the plasma in the disc and the differential frame-dragging are took into account. Therefore, the jet power is still a function of BH spin in the hybrid jet model. \citet{wu11} found that the jet efficiency of the hybrid model is consistent with those of MHD simulations very well. Following \citet{mei01}, the total jet power for the hybrid jet model is given by
  \be
   P_{\rm jet}={B_{\rm p}}^2 R^4 \Omega^2 /32c,
     \ee
where the poloidal magnetic field $B_{\rm p}\simeq gB_{\rm dynamo}$ \citep*[see][for more details]{mei01}. And the field-enhancing factor $g=\Omega/\Omega'$, where the disk angular velocity $\Omega$ is the sum of its angular velocity relative to the local metric $\Omega'$ plus the angular velocity of the metric itself in the Boyer-Lindquist frame $\omega\equiv -g_{\phi t}/g_{\phi \phi}$, i.e., $\Omega=\Omega'+\omega$. We set the jet formation region $R\simeq R_{\rm ms}$ \citep[e.g.,][]{nem07}, which is roughly consistent with the jet-launching regions in the numerical simulations \citep*[e.g.,][]{hir04}.

A fraction of the plasma in the accretion flow will be transferred into the jet along the poloidal magnetic field line that anchored in the ADAF. The material will be accelerated by the magnetic energy that extracted from the disk and the spinning BH. After the acceleration, most of the magnetic energy will be converted into the kinetic energy (e.g., $P_{\rm jet}=\Gamma(\Gamma-1)\dot{M}_{\rm jet}c^2$, $\Gamma$ is average bulk Lorentz factor and $\dot{M}_{\rm jet}$ is the mass-loss rate). The possible shock will occur as a result of the collision of shells with different velocities in the jet. To describe the jet radiation, we adopt the internal shock scenario that has been used to explain the broadband SED of XRBs, AGNs and afterglow of gamma-ray burst \citep*[e.g.,][]{pir99,yc05,wu07,nem12,xie16}. These shocks accelerate a fraction of the electrons, $\xi_{\rm e}$, into a power-law energy distribution with an index $p$. In this work, we fix $\xi_{\rm e}=0.01$ and allow the $p$ to be a free parameter that can be constrained from observations \citep*[e.g.,][]{yc05}. The energy density of accelerated electrons and amplified magnetic field are determined by two parameters, $\epsilon_{\rm e}$ and $\epsilon_{\rm B}$, which describe the fraction of the shock energy that goes into electrons and magnetic fields, respectively. Obviously, $\epsilon_{\rm e}$ and $\xi_{\rm e}$ are not independent. Only synchrotron emission is considered in calculation of the jet spectrum, where the synchrotron self-Compton in the jet is several orders of magnitude less than the synchrotron emission in X-ray band \citep*[see,][for more discussions]{wu07}. The jet inclination angle of $\sim15^{\rm o}$ is adopted for M87 \citep*[e.g.,][]{wz09}.

\section{Results}
  The millimeter bump of M87 as observed in the high-resolution quasi-simultaneous multi-waveband SED at a scale of 0.4 arcsec most possibly come from the innermost region of the ADAF close to the BH horizon \citep[]{feng16,li16}. The radio and X-ray emission should be dominated by the jet. Firstly, we adjust the wind parameter of ADAF, $s$ (control the accretion rate near the horizon), and BH spin, $a_{*}$, to fit the  millimeter bump and allow the calculated jet power (Equation 1) equal to the observed values as derived from the X-ray cavities. Then we model the radio and X-ray data with the jet model, where the mass-loss rate and velocity of jet are not independent at given jet power. In Figure 1, we present the fitting results, where the dotted line represents the ADAF spectrum, the dashed line represents the jet spectrum, and the solid line represents the total spectrum of ADAF and jet. From the top to bottom panels, we present the results for $a_{*}$=0.96, 0.98, and 0.992 for the jet power of $5\times10^{42}\rm erg/s$, $8\times10^{42}\rm erg/s$, and $1.5\times10^{43}\rm erg/s$ respectively (the model parameters are listed in Table 1). Therefore, we find the BH spin in M87 should be fast rotating with $a_{*}=0.98_{\rm -0.02}^{+0.012}$ for the constrained jet power of $8_{\rm -3}^{+7}\times10^{42}${\rm erg/s}. The magnetic field strength near the horizon is around 50-100 G in our model.

\begin{figure*}
\centering
\includegraphics[width=19cm,height=6cm]{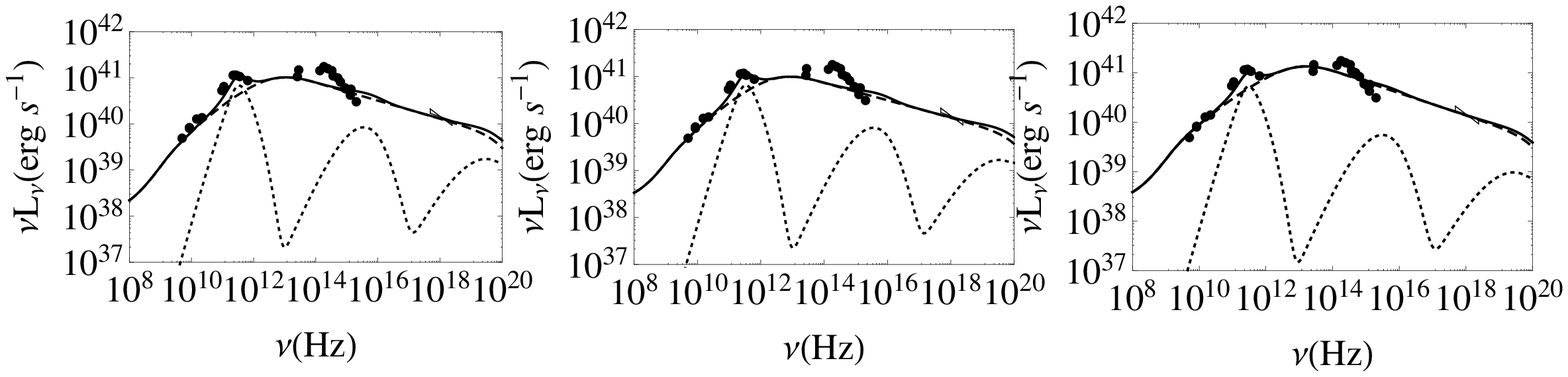}
\caption{ADAF-jet model results compared with the M87 0.4 arcsec aperture radius SEDs in quiescent state. The dotted lines represents the ADAF spectrum while the dashed lines show the jet spectrum. The solid line is the sum of ADAF and jet contribution. The ADAF parameters are $\alpha=0.3$, $\beta=0.5$, and $\delta=0.1$, respectively. From the left to right the black spin is 0.96, 0.98 and 0.992, respectively.
}
\label{fig:2009p2013}
\end{figure*}

\begin{table*}
\centering
\caption{Results from the hybrid model\label{tab:pks1424p240}.}
\begin{tabular}{ccccccc}
\hline \hline
$a_{*}$     & $R_{\rm ms} (\rm Rg)$   & $B_{\rm p (\rm G)}$  & $v_{\rm jet}$ ($c$) &   $\dot{M}_{\rm jet} (\rm M_\odot yr^{-1})$    &     $\eta$       & $P_{\rm jet} (\rm erg/s)$\\
\hline
0.96         & 1.84                  & 55            & 0.64       & $2.18\times10^{-4}$    & 0.089         & $5\times10^{42}$  \\
0.98         & 1.61                  & 78            & 0.7        & $2.33\times10^{-4}$    & 0.151         & $8\times10^{42}$  \\
0.992        & 1.41                  & 110           & 0.78       & $2.62\times10^{-4}$    & 0.343         & $1.5\times10^{43}$ \\

\hline \hline
\end{tabular}
\centering

 Note: The jet powers are calculated based on the hybrid model, and the parameters in the hybrid model are calculated based on the ADAF-jet model from the SED fitting. From left to right, the parameters are BH spin, innermost marginally stable orbit of the disk, amplified magnetic field, the speed of the jet, the mass-loss rate in the jet, radiative efficiency and calculated jet power, respectively.

\end{table*}

\section{Conclusion and Discussion}
The power of relativistic jet is correlated to the BH spin \citep[e.g.,][]{mck04,hir04,vill05,haw06,tch10,tch11,tch12a,tch12b,ny12}, which provide a method the constrain the BH spin parameter if the accretion rate can be measured. The millimeter/sub-millimeter polarimetry serve as an important tool for studying the hot plasma near a BH through the Faraday rotation of the polarized light \citep[]{kuo14,feng16,li16}. We fit the high-resolution multi-waveband SED of M87 with a coupled ADAF-jet model, where the millimeter bump is mainly contributed by the ADAF while the jet emission dominate at other waveband. We calculate the jet power using the best fit parameters, and find that the BH should be fast rotating with the dimensionless spin parameter $a_{*}=0.98_{\rm -0.02}^{+0.012}$ in M87 to reproduce the observed jet power of $8_{\rm -3}^{+7}\times10^{42}$ {\rm erg/s}.

The millimeter bump in M87 as recently observed by the Atacama Large Millimeter/submillimeter Array can be naturally modeled by the synchrotron emission of the thermal electrons in the ADAF, which is different from the prediction of the jet model. Therefore, it provide an opportunity to explore the accretion process near the BH horizon. In particular, \citet[]{feng16} and \citet[]{li16} both found that the RM predicted from the ADAF is roughly consistent with the observational values. It is still difficult to constrain the BH spin parameter from the modelling the SED of M87 due to there are some degeneracy in model parameters (wind parameter, $s$, magnetic parameter $\beta$, \citet[]{feng16}). The spin parameter can be better constrained from the jet model if the relativistic jet is indeed powered by the rotating BHs as suggested from MHD simulations and some observations. We find that the dimensionless BH spin parameter should be larger than 0.96 for the lower limit of jet power that derived from the X-ray cavities \citep[e.g.,][]{raf06,russ13}. In this work, we adopt several typical value of parameters (e.g., $\delta=0.1$, $\alpha=0.3$ and $\beta=0.5$). The larger value of $\delta$ will lead to lower accretion rate near the horizon to explain the observed millimeter bump and the BH need rotate faster to reproduce the observed jet power. The peak of synchrotron emission from the thermal electrons of ADAF will move to submillimeter waveband if $\delta$ value is too small, which is different from the observed millimeter bump. We adopt the equipartition case of $\beta=0.5$ in our calculations, where magnetic energy will become dominant if BH is fast spinning considering the possible amplification of the magnetic field by frame dragging effect. For weaker magnetic case (e.g., $\beta>0.5$), the BH need to rotate faster to explain the observed SED and jet power. We find our results are not sensitive to the viscosity parameter $\alpha$. Therefore, we suggest that BH should be fast rotating in M87 even consider the possible uncertainties.

Recent observations found that the jet may be structured not only in the radial direction but also in the transverse direction, where the jet may be composed by a fast spine flow and a surrounding slow layer \citep[e.g.,][]{gh05,tg08,xie12,mi15}. \citet{asa16} explored the jet structure of M87 by analyzing the high-resolution VSOP data, and they resolved the jet streamline into three ridgelines at the scale of milli arcseconds, where the outer streamline corresponds to the layer and the inner streamline corresponds to the spine. \citet{asa16} proposed that the inner ridgeline may originates from the spinning BH through the BZ process while the outer ridgelines may originate from the layer through the BP process from the disk. If the jet power measured from the X-ray cavity is mainly contributed by the outer slow layer, our BH spin estimation based on the hybrid model should be the upper limit. However, it is still difficult to estimate the jet power from the possible inner spine and outer layer respectively, where the plasma density, velocity and magnetic energy of the spine and layer cannot be well constrained from nowadays observations. \citet{tg08} modeled the TeV emission of M 87 using a structured jet with a fast spine surrounded by a slow layer, and found that most of the jet power may still be dominated by the spine (e.g., their Figure 1). If this is the case, our estimation on the BH spin should be acceptable since that the jet power contributed by the slow layer is less important. Furthermore, it should be difficult to understand why the large-scale jet was absent in most of other galaxies with similar BH mass and accretion rate as that in M87 if the relativistic jet power is normally contributed by the BP process (accretion mode is similar).  Future better measurements and more detailed theoretical calculation of the jet power for the putative fast spine and slow layer will help to further constrain the BH spin of M87.

 The fast variability of TeV photons from the center of M87 provide an another method to constrain the spin of its SMBH based on the escape of TeV photons from the radiation fields of ADAF near the horizon. Based on this method, \citet[]{li09} found that the BH spin should be $a_{*}\gtrsim0.8$ in M87 \citep[see also][]{wang08}, which is consistent with our result. Our result also strengthen the conclusion that the BHs in most radio galaxies may be rotating extremely fast \citep[eg.,][]{wu11}. \citet{dav11} also found a strong correlation between BH spin and BH mass, where the spin $j\sim 0$ for low-mass BHs with $M_{\rm BH} \sim 10^7 \msun$, while it is fast rotating with $j\sim 1$ for SMBHs with $M_{\rm BH} \sim 10^9 \msun$, based on the estimation of radiative efficiency for a sample of quasars \citep*[see][for a similar conclusion]{sik07,lag09,fan11}. The BH mass of M87 is larger than $10^9 \msun$ and may be indeed fast rotating.

In the coming few years, it is possible to resolve the strong field general relativistic signatures near the SMBH (e.g., Sgr A* and M87) with the Event Horizon Telescope (EHT), which is a project to assemble a Very Long Baseline Interferometry (VLBI) network of millimeter wavelength dishes.
Comparing no evident jet was found in Sgr A*, M87 provide a unique opportunity for understanding and testing the black hole physics and astrophysical processes in relativistic jet formation, collimation, and acceleration with the EHT. Our results show that the BH of M87 is fast spinning, which will predict much different BH shadow compared a non-spinning BH \citep[]{bar73,lu79,falc00,bro01,bro06a,bro06b,bro06,lu14,ak15,rd15,lu16}. We will explore the possible BH shadow in our following works.
\section*{Acknowledgements}

 This work is supported by the NSFC (11573009, 11133005 and 11622324).


\bsp	
\label{lastpage}

\end{document}